\documentclass[hidelinks,11pt]{article}  
\usepackage{amsthm,amssymb,amsmath}
\usepackage{graphicx}
\usepackage{booktabs}
\usepackage{mathtools}
\usepackage{setspace}
\usepackage{tabularx}
\usepackage{color}
\usepackage{epstopdf}
\usepackage[titletoc]{appendix}
\usepackage{cite}
\usepackage{natbib} 
\usepackage{float} 

\usepackage{multirow}
\usepackage{afterpage}
\usepackage{bm}
\usepackage{rotating}
\usepackage{centernot} 

\usepackage{enumerate} 
\usepackage[colorlinks=true]{hyperref}  
\hypersetup{
    linkcolor=blue,     
    citecolor=blue,      
    urlcolor=blue    
}

\usepackage{tikz}
\usetikzlibrary{calc}
\usetikzlibrary{decorations.pathreplacing}
\usetikzlibrary{matrix}

\newtheorem{assumption}{Assumption}[section]

\newtheorem{lemma}{Lemma}[section]
\newtheorem{theorem}[lemma]{Theorem}

\theoremstyle{definition}

\DeclareMathOperator*{\med}{med}
\DeclareMathOperator*{\sgn}{sgn}

\begin{document}

\title{Slope Consistency of Quasi-Maximum Likelihood  \smallskip\\ Estimator for Binary Choice Models%
\thanks{
We thank the 2024 Eminent Research Scholarship, University of Melbourne, for funding Yoosoon Chang and Joon Y. Park to visit the University of Melbourne.}
\medskip}
\date{\today}

\singlespacing
\author{Yoosoon Chang\thanks{Department of Economics,  Indiana University, \href{mailto:yoosoon@iu.edu}{yoosoon@iu.edu}}
\and Joon Y. Park\thanks{Department of Economics,  Indiana University, \href{mailto:joon@iu.edu}{joon@iu.edu}}
\and Guo Yan\thanks{Department of Economics, University of Melbourne, \href{mailto:yan.g@unimelb.edu.au}{yan.g@unimelb.edu.au}} }
\maketitle

\begin{abstract}
Although QMLE is generally inconsistent, logistic regression relying on the binary choice model (BCM) with logistic errors is widely used, especially in machine learning contexts with many covariates. This paper revisits the slope consistency of QMLE for BCMs. 
\cite{ruud-83} introduced a set of conditions under which QMLE may yield a constant multiple of the slope coefficient of BCMs asymptotically. However, he did not fully establish the slope consistency of QMLE, which requires the existence of a positive multiple of the true slope that maximizes the population QMLE likelihood over an appropriately restricted parameter space. 
We close this gap by providing a formal proof of slope consistency under the same set of conditions for BCMs identified as in \cite{manski-75,manski-85}. Our result implies that, under suitable conditions, logistic regression yields a consistent estimate of the slope coefficient for BCMs.

\end{abstract}

\vfill

\singlespacing
 \smallskip

\newpage
\renewcommand{\thefootnote}{\arabic{footnote}}
\setcounter{footnote}{0}
\onehalfspacing

\section{Introduction}

Logistic regression is widely used in empirical work and in machine learning to analyze binary outcomes, and is often applied as a quasi-maximum likelihood estimator (QMLE) for binary choice models (BCMs). When the error distribution in the underlying BCM is not logistic, the logit likelihood is misspecified, and the QMLE need not be consistent. Despite many alternative approaches for consistently estimating semiparametric BCMs,\footnote{E.g., \cite{powell-stock-stoker-89}, \cite{ichimura-93}, \cite{klein-spady-93}, \cite{ahn-ichimura-powell-ruud-18}, \cite{khan-lan-tamer-yao-24}, among others.} 
logistic regression remains common in applications, likely due to its computational simplicity and the availability of software packages.

This paper studies slope consistency of QMLE for BCMs. \cite{ruud-83} provided conditions under which QMLE for BCMs may asymptotically yield a slope vector proportional to the true slope. 
However, he did not fully establish the slope consistency of QMLE, which requires the existence of a positive multiple of the true slope that maximizes the population QMLE likelihood over an appropriately restricted parameter space. 
Without a careful argument establishing the existence of such a positive multiple, the proportionality constant need not be well-defined and, even when defined, could in principle be zero or negative, leading to incorrect conclusions (no effect or reversed sign). 
We close this gap by providing a formal proof of slope consistency under essentially the same conditions as in \cite{ruud-83} for BCMs identified as in \cite{manski-75,manski-85}. 

In the paper, we consider a binary outcome $Y$ given by the sign of $Y^\ast=\alpha_0+X'\beta_0-U$ with $X$ taking values in $\mathbb R^m$, and denote throughout by $\mathcal L$ the law or distribution.  
We impose (i) \emph{index dependence} $\mathcal L(U|X)=\mathcal L(U|V)$ with $V=\alpha_0+X'\beta_0$, and (ii) \emph{linearity in expectation} $\mathbb E(X|V)=aV+b$ for $a,b\in\mathbb R^m$.
\footnote{The index dependence condition is commonly imposed in the literature on consistent estimation of index models, including the papers cited in Footnote~1.} 
\cite{ruud-83} assumed linearity in expectation and independence of $X$ and $U$, which implies index dependence. Under (i)--(ii), together with necessary identification and regularity conditions (e.g., concavity and differentiability of the population log-likelihood), we establish slope consistency of the QMLE. Linearity in expectation is restrictive but holds, for instance, when $X$ is elliptically distributed, and can also be achieved by appropriate weighting; see \cite{ruud-86} and \cite{newey-ruud-94}.

Our results suggest that logistic regression can be used as a slope-consistent QMLE for an underlying BCM, provided that the index dependence and linearity-in-expectation conditions hold. 
This may provide some theoretical justification for the widespread use of logistic regression in machine learning for analyzing binary outcomes, as well as the popularity of logit and probit models in applied work.

The rest of the paper is organized as follows. Section~\ref{section-preliminaries} introduces the model and background theory with assumptions for identification and regularity in MLE. 
Section~\ref{section-slope-consistency} provides the slope consistency of the QMLE. Section \ref{section-concluding-remarks} concludes the paper. Mathematical proofs are in 
Appendix.

\section{Preliminaries}\label{section-preliminaries}

Consider the binary choice model (BCM) given by 
\begin{equation}\label{bcm-linear}
 Y = \sgn\,(Y^\ast)
\quad\mbox{with}\quad
 Y^\ast = \alpha_0 + X'\beta_0 - U,
\end{equation}
where $\sgn$ is the sign function defined as 
$\sgn\,(z) = \pm 1$ for $z\ge 0$ and $z<0$, respectively, $X$ is an $m$-dimensional vector of covariates, $\theta_0 = (\alpha_0,\beta_0')'$ is the true value of $\theta = (\alpha,\beta')'$, and $U$ is the error term.  
Let $\theta\in\Theta$, where $\Theta = \mathbb R^{m+1}$. 

Since we focus on the consistency of slope coefficient up to a positive scalar—referred to as slope consistency for simplicity—we impose conditions to ensure that $\theta_0$ is identified up to a positive scalar.

\begin{assumption}\label{med-zero} 
$\med\,(U|X) = 0$ almost surely in $\mathcal L(X)$. 
\end{assumption}

\begin{assumption}\label{pos-density}
(a) $X_m$ has a nonzero coefficient, and the distribution of $X_m$ conditional on $X_{-m}$ has everywhere positive Lebesgue density almost surely in $\mathcal L(X_{-m})$, where $X_{-m} = (X_1,\ldots,X_{m-1})'$, (b) $0 < \mathbb{P}\{Y=1|X \} <1$ almost surely in $\mathcal{L}(X)$, and (c) the support $\mathcal X$ of $\mathcal{L}(X)$ is not contained in any proper linear subspace of $\mathbb{R}^m$.
\end{assumption}

\noindent
Assumptions \ref{med-zero} and \ref{pos-density} ensure that $\theta_0 = (\alpha_0,\beta_0')'$ is identified up to multiplication by a positive scalar with $\beta_0\ne 0$; See, e.g., \cite{manski-75,manski-85}.  

We consider the maximum likelihood estimation of a possibly misspecified model, referred to as quasi-maximum likelihood (QML) estimation, which assumes that $U$ is independent of $X$ and has distribution function $F$. The QML estimator $\hat \theta=(\hat\alpha, \hat\beta')'$ is defined as the maximum of
\begin{equation}\label{qn} 
 Q_n(\theta) = \frac{1}{n} \sum_{i=1}^n \Big( 1\{Y_i=1\} \log F(\alpha+X_i'\beta) + 1\{Y_i=-1\}  \log \big( 1- F(\alpha+X_i'\beta)\big) \Big) 
\end{equation}
over $\Theta$. 
We define
\begin{equation}\label{q}
Q(\theta) =\mathbb{E}  \Big( 1\{Y=1\} \log F(\alpha+X'\beta) + 1\{Y=-1\}  \log \big(1- F(\alpha+X'\beta)\big) \Big) 
\end{equation}

\begin{assumption}\label{regularity}
(a) $Q$ has a maximum $\theta_\ast$ which is an interior point of $\Theta$, 
(b) $\mathbb E |\log F(\alpha+X'\beta) |, \mathbb E \big|\log \big(1- F(\alpha+X'\beta)\big)\big| <\infty$ for any $\theta\in \Theta$, and
(c) $F(\cdot)\in (0,1)$ on $\mathbb R$, and $\log F(\cdot), \log(1-F(\cdot))$ are strictly concave.  
\end{assumption}
\noindent
Assumption \ref{regularity} is standard. Part (a) assumes the existence of a pseudo-true value given by QMLE, which is commonly imposed in the existing work of QMLE, see, e.g., \cite{white-82}, \cite{gourieroux-monfort-trognon-84}. Part (b) ensures that $Q(\theta)$ is well-defined and $\hat Q(\theta) \to_p Q(\theta)$ for any $\theta \in\Theta$. Finally, Part (c) guarantees that $\theta_\ast$ is unique.

\begin{lemma}\label{plim-theta}
Let Assumption \ref{regularity} hold. Then $\hat \theta \to_p \theta_\ast$. 
\end{lemma}

\noindent  
Lemma~\ref{plim-theta} shows that $\hat \theta$ has a probability limit $\theta_\ast$. Lemma~\ref{plim-theta} follows immediately from Theorem 2.7 in \cite{newey-mcfadden-94}, since all conditions required in their theorem are trivially satisfied under our Assumption \ref{regularity}.

\begin{assumption}\label{differentiability}
The derivative $f$ of $F$ is well defined and continuously differentiable on $\mathbb R$.  
\end{assumption}
\noindent
Given Assumption \ref{regularity}, Assumption \ref{differentiability} implies that $\theta_\ast$ is uniquely defined as the solution to the first order condition (FOC) given by the derivative of $Q$. Both Assumption \ref{regularity}(c) and Assumption \ref{differentiability} are satisfied for normal and logistic distribution functions, which are most commonly used in practice.

Subsequently, we let Assumptions~\ref{regularity} and~\ref{differentiability} hold. 
Then $Q$ is strictly concave and twice continuously differentiable. Therefore, if we define
\begin{equation}\label{ldpm}
 \dot\ell_{+}(z) = \frac{f(z)}{ F(z)} 
\quad\text{and}\quad 
 \dot\ell_{-}(z) = \frac{f(z)}{1-F(z)},
\end{equation}
the probability limit $\theta_\ast = (\alpha_\ast,\beta_\ast')'$ of $\hat\theta$ is defined as the solution to the FOC $\dot Q(\theta) = 0$, where $\dot Q$ is the first-order derivative of $Q$ given by
\begin{equation}\label{qdot}
\dot Q(\theta) = \mathbb{E} \left(1\{Y=1\}\dot\ell_{+}(\alpha+X'\beta) - 1\{Y=-1\}\dot\ell_{-}(\alpha+X'\beta)\right) 
\begin{pmatrix}
1 \\ X \end{pmatrix}.
\end{equation}

\section{Slope Consistency}\label{section-slope-consistency}
Let 
\[
V = \alpha_0 + X'\beta_0,
\]
and consider the QMLE over a restricted set of parameters $\theta = (\alpha,\beta')'$ specified as
\begin{equation}\label{res}
 \begin{pmatrix} \alpha \\ \beta \end{pmatrix} 
 = c \begin{pmatrix} \alpha_0 \\ \beta_0 \end{pmatrix} 
 + \begin{pmatrix} r \\ 0 \end{pmatrix} 
\end{equation}
with some $c,r\in\mathbb R$. The resulting estimator will be referred to as the \emph{restricted} QMLE. Note that the restricted QMLE has only two parameters, while the unrestricted QMLE has $(m+1)$-parameters. To analyze the restricted QMLE, we rewrite $\dot Q(\theta)$ in \eqref{qdot} as
\begin{equation}\label{nsc}
\dot Q(c,r) = \mathbb E\left( 1\{U\le V\}\dot\ell_{+}(cV+r) - 1\{U>V\}\dot\ell_{-}(cV+r)\right) 
 \begin{pmatrix} 1 \\ X \end{pmatrix}
\end{equation}
with the $(m+1)$-dimensional parameter $\theta = (\alpha,\beta')'$ being restricted to the two dimensional parameter $(c,r)$ introduced in \eqref{res}. Then we may easily deduce that

\begin{lemma}\label{iff-consistency}
Let Assumptions \ref{med-zero}, \ref{pos-density}, \ref{regularity} and \ref{differentiability} hold. Then the QMLE is slope consistent if and only if $\dot Q(c,r) = 0$ has a solution $(c_\ast,r_\ast)$ such that $c_\ast >0$ and $r_\ast \in \mathbb R$.
\end{lemma}

Now we introduce a set of sufficient conditions to ensure that $\dot Q(c,r) = 0$ has a solution $(c_\ast,r_\ast)$ such that $c_\ast >0$ and $r_\ast \in \mathbb R$.

\begin{assumption}\label{id-dep} $\mathcal L(U|X) = \mathcal L(U|V)$.
\end{assumption}

\noindent
Assumption \ref{id-dep} requires that the error distribution depends on $X$ only through the index $V$. Such an \emph{index dependence} condition for the error distribution is often used in the literature.\footnote{
See, e.g., \cite{klein-spady-93} for the BCM, and \cite{ichimura-93} for a class of single index models. 
}

\begin{assumption}\label{ld-mean} $\mathbb E(X|V) = aV + b$ for some $a,b \in \mathbb R^m$.
\end{assumption}

\noindent
Assumption \ref{ld-mean} requires that the conditional mean of $X$ given $V$ is a linear function of $V$, which will be referred to simply as the \emph{linearity in expectation} condition. This condition is restrictive, but it holds, for instance, if $X$ has an elliptical distribution. 
We may also weight observations on $X$ appropriately so that the distribution of the weighted observations can be regarded as satisfying this condition; see, e.g., \cite{ruud-86} and \cite{newey-ruud-94}.\footnote{
Specifically, in place of the original observations $\{x_i\}_{i=1}^n$ on $X$, they use the observations weighted by $w_i = \sigma(x_i)/\hat\tau(x_i)$ for $i=1,\ldots,n$, where $\sigma$ is the standard multivariate normal density satisfying Assumption~\ref{ld-mean}, and $\hat\tau$ is a local kernel density estimator of $X$. The weighted observations may be approximately regarded drawn from a distribution with density $\sigma$. 
}

In what follows, under Assumptions \ref{med-zero}, \ref{pos-density}, \ref{regularity} and \ref{differentiability}, we establish that Assumptions \ref{id-dep} and \ref{ld-mean} indeed ensure the existence of solution $(c_\ast,r_\ast)$ with $c_\ast > 0$ to $\dot Q(c,r) = 0$. It is clear that $\dot Q(c,r) = 0$ may have such a solution without Assumptions \ref{id-dep} and \ref{ld-mean}, so they need not be necessary.

Let 
\[
\Pi(v) = \mathbb P\big\{Y=1\big|V=v\big\} = \mathbb P\big\{U\leq V\big|V=v\big\} 
\]
for $v\in\mathbb R$. Due to Assumption \ref{med-zero}, we have $\Pi(v)\ge 1/2$ if $v\ge 0$, and $\Pi(v) < 1/2$ if $v<0$. Under Assumptions~\ref{id-dep} and~\ref{ld-mean}, $\dot Q(c,r)$ defined in \eqref{nsc} becomes\footnote{
To see this, first use the law of iterated expectation (LIE) by taking the conditional expectation on $X$, where we simplify $\mathbb E  ( 1\{U\le V\} |X) = \Pi(V), \mathbb E  ( 1\{U> V\} |X) = 1-\Pi(V)$ under Assumption~\ref{id-dep}. Second, use LIE by taking the conditional expectation on $V$ and note $\mathbb E(X|V) = a V+ b$ by Assumption~\ref{ld-mean}. 
}
\begin{equation}\label{mpo}
 \dot Q(c,r) = \mathbb E\left(\Pi(V)\dot\ell_{+}(cV + r) - (1-\Pi(V))\dot\ell_{-}(cV+r)\right) 
 \begin{pmatrix} 1 \\ a V+ b \end{pmatrix},
\end{equation}
and consequently, the system of $(m+1)$-equations $\dot Q(c,r) = 0$ effectively reduces to the system of two equations given by
\begin{align}\label{foc}
 \dot Q_\bullet(c,r) 
 &= \mathbb E\left(\Pi(V)\dot\ell_{+}(cV+r) - (1-\Pi(V))\dot\ell_{-}(cV+r)\right) 
 \begin{pmatrix} 1 \\ V \end{pmatrix} = 0
\end{align}
with two unknowns $(c,r)$.

Slope consistency of the QMLE requires that the FOC $\dot Q_\bullet(c,r)=0$ has a solution $(c_\ast,r_\ast)$ with $c_\ast>0$. 
However, the existence of such a solution is not automatic. 
Even if Assumption~\ref{regularity} ensures that the population likelihood for the unrestricted QMLE has a unique maximizer, it does \emph{not} mean that the population likelihood for the restricted QMLE also has a maximizer. 
Moreover, even if the FOC admits a solution, the associated $c_\ast$ need not be positive. 

In fact, \cite{ruud-83} shows neither that the FOC exists a solution $(c_\ast,r_\ast)$ nor that $c_\ast>0$; instead, he assumes that the restricted likelihood (which is essentially identical to ours) is maximized at a point satisfying the FOC. 
\cite{li-duan-89} consider a more general class of models and impose an additional high-level condition that guarantees, for BCMs, existence of a solution $c_\ast\in\mathbb R$, but not necessarily with $c_\ast>0$. See also \cite{powell-94} for related discussion. 
We address these issues by establishing the following lemma, which is the key technical contribution of the paper.

\begin{lemma}\label{cast-positive}
Let Assumptions \ref{med-zero}, \ref{pos-density}, \ref{regularity} and \ref{differentiability} hold. Then $\dot Q_\bullet(c,r) = 0$ has a solution $(c_\ast,r_\ast)$ such that $c_\ast >0$ and $r_\ast \in \mathbb R$. 
\end{lemma}

\noindent
The proofs for Lemma~\ref{cast-positive} and Theorem~\ref{main} are provided in 
Appendix. 

\begin{theorem}\label{main}
Let Assumptions \ref{med-zero}, \ref{pos-density}, \ref{regularity}, \ref{differentiability}, \ref{id-dep} and \ref{ld-mean} hold. Then $\dot Q_\bullet(c,r) = 0$ has a unique solution $(c_\ast,r_\ast)$ for $c_\ast>0$ and $r_\ast\in\mathbb R$. 
Moreover, $\hat\alpha \to_p c_\ast\alpha_0 + r_\ast$ and $\hat\beta \to_p c_\ast \beta_0$ as $n\to\infty$. 
\end{theorem}

\noindent
Theorem \ref{main}, which follows immediately from Lemma \ref{cast-positive} under additional conditions in Assumptions \ref{id-dep} and \ref{ld-mean}, establishes slope consistency of QMLE for binary choice models.

Under suitable regularity conditions, $\sqrt{n}(\hat\theta-\theta_\ast) = -\big[\ddot Q_n(\theta_\ast)\big]^{-1}\big[\sqrt{n}\dot Q_n(\theta_\ast)\big] + o_p(1)$ and $\sqrt{n}\dot Q_n(\theta_\ast)$ has a normal limit distribution, where $\dot Q_n$ and $\ddot Q_n$ are the first and second derivatives, respectively, of $Q_n$ defined in \eqref{qn}. 
In this case, inference for $\beta_\ast$ in $\theta_\ast = (\alpha_\ast,\beta_\ast')'$ can be conducted using standard QMLE theory with a robust (sandwich) variance; see \cite{white-82}.
Since $\beta_\ast = c_\ast\beta_0$, we can test scale-invariant hypotheses about $\beta_0$, such as $\beta_{j,0} = 0$ and $\beta_{j,0} = \beta_{k,0}$ for $j\ne k$, among many others. 
Such scale-invariant hypotheses are natural: in our setup $\beta_0$ is identified only up to a positive scale, and even in BCMs with a scale normalization, the normalization is typically a convention that fixes the scale of latent utility and carries no economic content.

\section{Concluding Remarks}\label{section-concluding-remarks}

This note establishes slope consistency of the QMLE for BCMs under index dependence and linearity in expectation, together with standard regularity conditions. 
These results provide conditions under which logistic regression is slope-consistent as a QMLE for an underlying BCM, and may provide some theoretical justification for the popularity of logit/probit models in applied work. 
The main substantive requirement is linearity in expectation, which holds when covariates are elliptically distributed and may also be satisfied by reweighting observations; see \cite{ruud-86} and \cite{newey-ruud-94}. 
We focus on slope consistency because empirical work often relies on the relative magnitudes of slope coefficients to assess the role of covariates in latent utilities, and the intercept can be estimated separately once the slope is consistently estimated.

\appendix

\section*{Appendix: Mathematical Proofs}
\renewcommand{\thesubsection}{A.\arabic{subsection}}

 \subsection{Proof of Lemma \ref{cast-positive}}
The proof of Lemma~3.2 
will be done in three steps. We let
\begin{align*}
	\phi(c,r) &= \mathbb{E}\Big(1\{U\le V\}\dot\ell_{+}(cV+r)-1\{U>V\}\dot\ell_{-}(cV+r)\Big) \\
	&= \mathbb{E}\Big(\Pi(V)\dot\ell_{+}(cV+r)-\big(1-\Pi(V)\big)\dot\ell_{-}(cV+r)\Big) 
\end{align*}
and
\begin{align*}
	\psi(c,r) &= \mathbb{E}\Big(1\{U\le V\}\dot\ell_{+}(cV+r)-1\{U > V\}\dot\ell_{-}(cV+r)\Big)V \\
	&= \mathbb{E}\Big(\Pi(V)\dot\ell_{+}(cV+r)-\big(1-\Pi(V)\big)\dot\ell_{-}(cV+r)\Big)V, 
\end{align*}
and show that the two equations given by $\phi(c,r) = 0$ and $\psi(c,r) = 0$ have a solution $(c_\ast, r_\ast)$ with $c_\ast > 0$ and $r_\ast\in\mathbb R$. In the first step, we show that there exists a unique $r(c) \in \mathbb R$ such that $\phi(c,r(c))=0$ for any $c\ge 0$. Then we let
\[
\psi(c) = \psi\big(c,r(c)\big),
\]
and show that $\psi(0) >0$ and $\psi(c) < 0$ for some $c>0$ in the second and third steps, respectively. The first and second steps are straightforward, while the third step is more involved. Throughout the proofs, we repeatedly use the fact that $\dot\ell_{+}(z) \big(\dot\ell_{-}(z)\big)$ is continuous, and strictly decreasing (increasing) to zero (from zero) on $\mathbb R$ as $z\to\infty$ with intersection at $z=0$ with their value $2f(0)$.\footnote{
	Note that $\dot \ell_{+}$ is strictly decreasing and $\dot \ell_{-}$ is strictly increasing on $\mathbb R$, since $\log F, \log (1-F)$ are strictly concave. 
}

\paragraph{Step 1:} In this step, we fix $c\ge 0$ arbitrarily, and show that there exists a unique $r(c) \in \mathbb R$ such that $\phi(c,r(c))=0$. Note that
\[
\lim_{r\to-\infty}\phi(c,r) \in (0,\infty]
\quad\mbox{and}\quad
\lim_{r\to\infty}\phi(c,r) \in [-\infty,0),
\]
and that $\phi(c,r)$ is continuous and strictly decreasing in $r\in\mathbb R$ for any given $c\ge 0$. Therefore, for any $c\ge 0$, there exists a unique $r(c)\in\mathbb R$ such that $\phi(c,r(c))=0$ by the intermediate value theorem. In particular, $r(0)$ is defined as
\begin{equation}\label{pf-psi-c-zero-aux-1}
	\phi(0,r(0)) = \dot\ell_{+}\big(r(0)\big)\mathbb P\{U\le V\} 
	- \dot\ell_{-}\big(r(0)\big)\mathbb P\{U>V\} = 0,
\end{equation}
which will be used subsequently.

\paragraph{Step 2:} In this step, we show that $\psi(0)>0$. This follows immediately, since
\begin{align}\label{pf-psi-c-zero-aux-2}
	\psi(0) &= \dot\ell_{+}\big(r(0)\big)\mathbb E\big(V1\{U\le V\}\big)
	- \dot\ell_{-}\big(r(0)\big)\mathbb E\big(V1\{U>V\}\big) \notag \\
	&= \dot\ell_{+}\big(r(0)\big)\mathbb E\big(V\big|U\le V\big)\mathbb P\{U\le V\}
	- \dot\ell_{-}\big(r(0)\big)\mathbb E\big(V\big|U > V\big)\mathbb P\{U > V\} \notag \\
	& = \dot\ell_{+}\big(r(0)\big)\Big(\mathbb E\big(V\big|U\le V\big) 
	- \mathbb E\big(V\big|U>V\big)\Big)\mathbb P\{U\le V\} > 0
\end{align}
under \eqref{pf-psi-c-zero-aux-1}. This was to be shown. Note that $\mathbb P\{U\le V\} > 0$ by Assumption~2.2(b), 
and that $\dot\ell_{+}(r(0)) >0$ since $\dot\ell_{+}$ is strictly decreasing with $\dot\ell_{+}(\infty)=0$ on $\mathbb R$.

\paragraph{Step 3:} In this step, we show that
\begin{equation}\label{step-3}
	\psi(c) < 0
\end{equation}
for some $c>0$. Here we define $\psi(c)$ equivalently as 
\begin{align*}
	\psi(c) &= \mathbb{E}\Big(\Pi(V)\dot\ell_{+}(cV+r(c))
	-\big(1-\Pi(V)\big)\dot\ell_{-}(cV+r(c))\Big)\big(cV+r(c)\big) \\ 
	&= \mathbb{E}\Big(\Pi(V)\dot\ell_{+}\big(c(V+s(c))\big)
	-\big(1-\Pi(V)\big)\dot\ell_{-}\big(c(V+s(c))\big)\Big)\big(c(V+s(c))\big) 
\end{align*}
with $s(c) = r(c)/c$ for $c>0$. Under this definition, we may easily see why \eqref{step-3} should hold. For any small $\delta>0$, there exists a large enough $c>0$ such that
\begin{align*}
	& \Pi(v)\dot\ell_{+}\big(c(v+s(c))\big) - \big(1-\Pi(v)\big)\dot\ell_{-}\big(c(v+s(c))\big) \\
	& \quad  \approx 
	\begin{cases}
		- \big(1-\Pi(v)\big)\dot\ell_{-}\big(c(v+s(c))\big) < 0 & \text{when}\ v+s(c)>\delta  \\
		\Pi(v)\dot\ell_{+}\big(c(v+s(c))\big) > 0 & \text{when}\ v+s(c)<-\delta
	\end{cases},
\end{align*}
from which it follows that
\[
\Big(\Pi(v)\dot\ell_{+}\big(c(v+s(c))\big) 
-\big(1-\Pi(v)\big)\dot\ell_{-}\big(c(v+s(c))\big)\Big)(v+s(c)) < 0
\]
for all $v\in\mathbb R$.

In the rest of the proof, we will show it more rigorously. Let $\{c_n\}$ be a sequence of numbers such that $c_n>0$ for all $n$ and $c_n \to\infty$ as $n\to\infty$, and define $s_n = s(c_n)$ for all $n$. Then it suffices to show that
\begin{align}\label{pf-psi-cinf-positive-aux-main}
	\liminf_{n\to \infty} \mathbb E  \Big( \Pi(V) \dot\ell_{+}( c_n (V+ s_n)  ) -   \big( 1- \Pi(V) \big) \dot\ell_{-}( c_n (V+ s_n)   ) \Big) (V + s_n)  < 0
\end{align}
as $n\to\infty$, to establish \eqref{step-3}. We may assume without loss of generality, by taking a subsequence if necessary, that the sequence $\{s_n\}$ is monotonic. Since $\{s_n\}$ is monotonic by construction, we have either (i) $s_n \to \pm \infty$ or (ii) $s_n\to \bar s\in \mathbb R$ as $n\to\infty$, depending upon whether $\{s_n\}$ is unbounded or bounded.

Since $f$ is a density function, we have
\[
f(z) = o\left(\frac{1}{|z|}\right)
\]
as $z\to \pm \infty$, from which it follows that
\[
\lim_{z\to \infty} \frac{\dot\ell_{+}(z)}{1/|z|} = \lim_{z\to \infty} \frac{|z| f(z)}{F(z)} = 0, 
\quad  \lim_{z\to -\infty} \frac{\dot\ell_{-}(-z)}{1/|z|} =  \lim_{z\to -\infty} \frac{ |z|f(z)}{1-F(z)}  =0,
\]
and that
\[
\sup_{z\geq 0} \frac{\dot\ell_{+}(z)}{1/|z|} <\infty, \quad \sup_{z\leq 0} \frac{\dot\ell_{-}(z)}{1/|z|} <\infty. 
\]
For any $\delta>0$ small, we therefore have 
\begin{align*}
	\Big| \mathbb E 1\{V+s_n>\delta\} \Pi(V) \dot\ell_{+} \big( c_n(V+s_n) \big) c_n (V+s_n) \Big| &\leq \sup_{z\geq 0} \frac{\dot\ell_{+}(z)}{1/|z|} <\infty \\
	\Big| \mathbb E 1\{V+s_n<-\delta\} \big(1- \Pi(V) \big) \dot\ell_{-} \big( c_n(V+s_n) \big) c_n (V+s_n) \Big| &\leq \sup_{z\leq 0} \frac{\dot\ell_{-}(z)}{1/|z|} <\infty, 
\end{align*}
from which we may easily deduce that
\begin{align*}
	\mathbb E 1\{V+s_n>\delta\}\Pi(V)\dot\ell_{+}\big(c_n(V+s_n)\big)(V+s_n) &\to 0 \\
	\mathbb E 1\{V+s_n<-\delta\}\big(1-\Pi(V)\big)\dot\ell_{-}\big(c_n(V+s_n)\big)(V+s_n) &\to 0
\end{align*}
as $n\to\infty$.

It follows that 
\begin{align}\label{decompn-psi-aux-1}
	& \mathbb E 1\{V+s_n>\delta\}  \Big( \Pi(V) \dot\ell_{+} \big( c_n(V+s_n) \big) - \big( 1- \Pi(V) \big)  \dot\ell_{-} \big( c_n(V+s_n) \big) \Big) (V+s_n) \notag \\
	& = o(1) - \mathbb E 1\{V+s_n>\delta\} \big( 1- \Pi(V) \big) \dot\ell_{-} \big( c_n(V+s_n) \big) (V+s_n)  \notag \\
	& < o(1) - \mathbb E 1\{V+s_n>\delta\} \big( 1- \Pi(V) \big) \dot\ell_{-}( c_n \delta ) \delta
\end{align}
and 
\begin{align}\label{decompn-psi-aux-2}
	& \mathbb E 1\{V+s_n<-\delta\}  \Big( \Pi(V) \dot\ell_{+} \big( c_n(V+s_n) \big) - \big( 1- \Pi(V) \big)  \dot\ell_{-} \big( c_n(V+s_n) \big) \Big) (V+s_n) \notag  \\
	& = o(1) + \mathbb E 1\{V+s_n<-\delta\} \Pi(V)  \dot\ell_{+} \big( c_n(V+s_n) \big) (V+s_n)  \notag  \\
	& < o(1) - \mathbb E 1\{V+s_n<-\delta\} \Pi(V)  \dot\ell_{+} (-c_n\delta) \delta
\end{align}
for all large $n$. Furthermore, we have 
\begin{align*} 
	\Big|\mathbb E1\{-\delta\!<\!V\!+\!s_n\!<\!0\}\big( 1\!-\!\Pi(V) \big) \dot\ell_{-}\big( c_n(V\!+\!s_n) \big) (V\!+\!s_n) \Big| &\leq \mathbb P\{-\delta\!<\!V\!+\!s_n\!<\! 0\}\dot\ell_{-}(0)\delta \\
	\Big|\mathbb E1\{0\!<\!V\!+\!s_n\!<\!\delta \} \Pi(V) \dot\ell_{+}\big( c_n(V\!+\!s_n) \big) (V\!+\!s_n) \Big| &\leq \mathbb P\{0\!<\!V\!+\!s_n\!<\!\delta \}\dot\ell_{+}(0) \delta,
\end{align*}
from which we may deduce that
\begin{align}\label{decompn-psi-aux-3}
	& \mathbb E 1\{-\delta<V\!+\!s_n<\delta\} \Big(\Pi(V)\dot\ell_{+}\big( c_n(V\!+\!s_n) \big) - \big( 1- \Pi(V) \big) \dot\ell_{-} \big( c_n(V\!+\!s_n) \big) \Big) (V\!+\!s_n) \notag \\
	& \leq \Big( \dot\ell_{-}(0) \mathbb P\{-\delta <V\!+\!s_n < 0\} +  \dot\ell_{+}(0) \mathbb P\{0 <V\!+\!s_n < \delta \} \Big) \delta \notag  \\
	& \quad + \mathbb E 1\{-\delta <V\!+\!s_n <0\} \Pi(V) \dot\ell_{+} \big( c_n(V\!+\!s_n) \big) (V\!+\!s_n) \notag  \\
	& \quad -\mathbb E 1\{0 <V\!+\!s_n< \delta\} \big( 1-\Pi(V) \big)  \dot\ell_{-} \big( c_n(V\!+\!s_n) \big) (V\!+\!s_n) \notag \\
	& < \Big( \dot\ell_{-}(0) \mathbb P\{-\delta <V\!+\!s_n < 0\} +  \dot\ell_{+}(0) \mathbb P\{0 <V\!+\!s_n < \delta \}  \Big) \delta
\end{align}
for all large $n$. Consequently, it follows from \eqref{decompn-psi-aux-1}, \eqref{decompn-psi-aux-2} and \eqref{decompn-psi-aux-3} that
\begin{align}\label{decompn-psi-aux-main}
	& \mathbb E \Big( \Pi(V) \dot\ell_{+} \big( c_n(V\!+\!s_n) \big) - \big( 1- \Pi(V) \big)  \dot\ell_{-} \big( c_n(V\!+\!s_n) \big) \Big) (V\!+\!s_n)  \notag \\
	& < o(1) - \dot\ell_{-}( c_n \delta ) \delta  \mathbb E 1\{V\!+\!s_n>\delta\} \big( 1- \Pi(V) \big)  - \dot\ell_{+} (-c_n\delta) \delta  \mathbb E 1\{V\!+\!s_n<-\delta\} \Pi(V) \notag \\  
	& \quad + \bigg( \dot\ell_{-}(0) \mathbb P\{-\delta <V\!+\!s_n < 0\} +  \dot\ell_{+}(0)  \mathbb P\{0 < V\!+\!s_n < \delta \}  \Big) \delta  \notag \\ 
	& < o(1) - \delta \Big( \dot\ell_{-}(0) \mathbb E 1\{V\!+\!s_n > \delta\} \big( 1- \Pi(V) \big)  + \dot\ell_{+} (0) \mathbb E 1\{V\!+\!s_n < -\delta\} \Pi(V) 
	\notag  \\
	& \qquad \qquad \qquad - \big(\dot\ell_{-}(0) + \dot\ell_{+}(0) \big) \mathbb P\{-\delta <V\!+\!s_n < \delta \} \bigg)
\end{align} 
for all large $n$.

If $s_n \to \pm\infty$, then $\mathbb P\{-\delta <V\!+\!s_n <\delta \} \to 0$. In this case, \eqref{decompn-psi-aux-main} yields
\begin{align*}
	& \liminf_{n\to \infty} \mathbb E \Big( \Pi(V) \dot\ell_{+} \big( c_n(V\!+\!s_n) \big) - \big( 1- \Pi(V) \big)  \dot\ell_{-} \big( c_n(V\!+\!s_n) \big) \Big) (V\!+\!s_n) \\
	& \leq \liminf_{n\to\infty} \bigg[ -\delta \Big( \dot\ell_{-}(0) \mathbb E 1\{V\!+\!s_n> \delta\} \big( 1- \Pi(V) \big)  + \dot\ell_{+} (0) \mathbb E 1\{V\!+\!s_n < -\delta\} \Pi(V) \Big) \bigg] < 0,
\end{align*}
since $\dot\ell_{+}(0),\dot\ell_{-}(0)>0$ and $\mathbb E\Pi(V)\in(0,1)$, and therefore, \eqref{pf-psi-cinf-positive-aux-main} holds. On the other hand, in case $s_n \to \bar s\in \mathbb R$, \eqref{decompn-psi-aux-main} yields
\begin{align*}
	& \liminf_{n\to \infty}\mathbb E \Big( \Pi(V) \dot\ell_{+} \big( c_n(V\!+\!s_n) \big) - \big( 1- \Pi(V) \big)  \dot\ell_{-} \big( c_n(V\!+\!s_n) \big) \Big) (V\!+\!s_n)  \\
	& \leq - \delta \Big( \dot\ell_{-}(0) \mathbb E 1\{V\!+\!\bar s> \delta\} \big( 1- \Pi(V) \big)  + \dot\ell_{+} (0) \mathbb E 1\{V\!+\!\bar s< -\delta\} \Pi(V)  \\
	& \qquad \qquad - \big(\dot\ell_{-}(0) + \dot\ell_{+}(0) \big) \mathbb P\{ |V\!+\!\bar s| < \delta \} \Big)  < 0
\end{align*}
for all $\delta>0$ small enough, since $\mathbb P\{ |V+\bar s| < \delta \} \to 0$ as $\delta \to 0$ due to Assumption~2.2 
which implies that $V$ has a positive Lebesgue density on $\mathbb R$. Therefore, it follows that \eqref{pf-psi-cinf-positive-aux-main} holds also in this case.

With Steps 1, 2, and 3 established, we are now ready to prove Lemma~3.2. 
Since $\phi(c,r)$ is continuous in $(c,r)$, we may readily deduce that $r(c)$ is continuous in $c$. Likewise, $\psi(c,r)$ is continuous in $(r,c)$, from which it follows that $\psi(c)$ is continuous in $c>0$. Therefore, Step 2 and Step 3 imply that $\psi(c) = 0$ has a solution $c_\ast>0$ by the intermediate value theorem. Once $c_\ast>0$ is obtained, we may let $r_\ast\in \mathbb R$ be given by $r_\ast = r(c_\ast)$, where $r(c)$ is defined in Step 1. It is clear that $(c_\ast,r_\ast)$ is a solution to $\dot Q_\bullet(c,r) = 0$, as was to be shown. This completes the proof.

 \subsection{Proof of Theorem \ref{main}}
Let $(c_\ast,r_\ast)$ be a solution to $\dot Q_\bullet(c,r) = 0$ with $c_\ast>0$ and $r_\ast \in \mathbb R$, whose existence is guaranteed by Lemma~3.2 and Assumptions~2.1--2.4. 
By Assumptions~3.1--3.2, $(c_\ast,r_\ast)$ is also  a solution to $\dot Q(c,r) = 0$, as shown in Eq.~(8). 
Thus, $(c_\ast\alpha_0 + r_\ast, c_\ast \beta_0')'$ is a solution to $\dot Q(\theta) = 0$. 
Since the probability limit $\theta_\ast$ of $\hat\theta$ in Lemma~2.1 is uniquely chacterized by $\dot Q(\theta_\ast) = 0$ under Assumptions~2.3--2.4, $\theta_\ast = (c_\ast\alpha_0 + r_\ast, c_\ast \beta_0')'$. Thus, $\hat\alpha\to_p c_\ast \alpha_0 +r_\ast$ and $\hat\beta \to_p c_\ast \beta_0$ by Lemma~2.1.

\bibliographystyle{econometrica}
\bibliography{bcm-qmle}

\end{document}